\documentclass[aps,amsmath,amssymb,floatfix,twocolumn]{revtex4}

\usepackage{bm}
\usepackage{graphicx}
\usepackage{amsbsy}
\usepackage{dcolumn}
\usepackage{color}

\newcommand{\be}{\begin{equation}}
\newcommand{\ee}{\end{equation}}
\newcommand{\bea}{\begin{eqnarray}}
\newcommand{\eea}{\end{eqnarray}}

\begin{document}

\title{Equation of state of nuclear and neutron matter at \\ 
       third-order in perturbation theory from chiral EFT}
\author{J.\ W.\ Holt$^1$ and N.\ Kaiser$^2$}
\affiliation{$^1$Cyclotron Institute and Department of Physics and Astronomy, Texas 
A\&M University, College Station, TX}  
\affiliation{$^2$Physik Department, Technische Universit\"{a}t M\"{u}nchen, Garching, 
Germany}

\begin{abstract}
We compute from chiral two- and three-nucleon interactions the energy per particle of symmetric nuclear 
matter and pure neutron matter at third-order in perturbation theory 
including self-consistent second-order single-particle energies.
Particular attention is paid to the third-order particle-hole ring-diagram, which is often neglected in 
microscopic calculations of the equation of state. We provide semi-analytic expressions for the 
direct terms from central and tensor model-type interactions that are useful as theoretical benchmarks.
We investigate uncertainties arising from the order-by-order convergence in both 
many-body perturbation theory and the chiral expansion. Including also variations in the resolution 
scale at which nuclear forces are resolved, we provide new error bands on the equation of state,
the isospin-asymmetry energy, and its slope parameter. We find in particular that the inclusion of 
third-order diagrams reduces the theoretical uncertainty at low densities, while in general
the largest error arises from omitted higher-order terms in the chiral expansion of the nuclear forces.
\end{abstract}

\maketitle

\section{Introduction}

The nuclear isospin-asymmetry energy $S(\rho)$, defined as the difference 
between the energy per particle of homogeneous neutron matter and symmetric nuclear matter at a 
given density $\rho$, offers important links between the properties of terrestrial nuclei and extreme 
astrophysical systems such as core-collapse 
supernovae, neutron stars, and neutron star mergers \cite{lattimer00,steiner05,lattimer07}. 
Clarifying the experimental and theoretical uncertainties on the isospin-asymmetry 
energy is therefore an important objective in contemporary low-energy nuclear physics.
Intermediate-energy heavy-ion collision experiments 
\cite{danielewicz02,natowitz10,qin12,horowitz14} can access the equation of state of nuclear 
matter at supra-saturation densities, and modern theoretical methods such as chiral 
effective field theory (EFT) \cite{weinberg79,epelbaum09rmp,machleidt11} and the renormalization group 
\cite{bogner03,bogner10} allow for nuclear matter calculations with reliable uncertainty estimates
up to roughly $\rho \simeq 2 \rho_0$, where $\rho_0\simeq 0.16\,\text{fm}^{-3}$ is the saturation density. 
These estimates are achieved through systematic studies \cite{sammarruca15,epelbaum15} of the 
order-by-order convergence in the chiral power counting together with 
variations in the resolution scale \cite{bogner05} and low-energy constants \cite{hebeler10} in the
chiral two-body ($2N$) and three-body ($3N$) potentials. In addition, the comparison of different 
perturbative \cite{fritsch05,hebeler11,coraggio13,coraggio14} and nonperturbative methods 
\cite{epelbaum09epja,gezerlis13,kohno13,roggero14,carbone14,hagen14,wlazlowski14} for 
solving the nuclear many-body problem starting from identical Hamiltonians can give insight 
into additional sources of error. In all cases, both theoretical and experimental uncertainty bands grow 
rapidly with the density beyond $\rho = \rho_0$. In the vicinity of nuclear matter saturation density,
microscopic calculations \cite{hebeler10prl,gandolfi12} of the isospin-asymmetry energy $S(\rho)$ and its 
slope parameter $L$ tend to lie just outside of the experimental band \cite{lattimer13,lattimer14}.
More detailed investigations of theoretical uncertainties may therefore shed light on this discrepancy.

In the present work, we explore the role of third-order perturbative contributions to the 
nuclear and neutron matter equations of state as well as the effects of self-consistent 
second-order single-particle energies. Numerous works \cite{hebeler11,coraggio13,tews16,drischler16} 
have studied the
importance of particle-particle and hole-hole ladder diagrams at third order, but 
the third-order particle-hole diagram is often neglected due to its more complicated momentum and 
spin recouplings when expressed in terms of partial waves, which significantly increase the computational 
cost. In Ref.\ \cite{coraggio14} it was found that the third-order particle-hole
diagram gives a contribution on the order of $|E^{(3)}_{ph}(\rho_0)| \sim (1-2)$\,MeV in symmetric 
nuclear matter when computed from coarse-resolution chiral $N\!N$-potentials with momentum-space 
cutoffs $\Lambda \lesssim 500$\,MeV. Third-order contributions are
expected to be less important in neutron matter, which generally exhibits faster convergence 
\cite{krueger13,rrapaj16} in many-body perturbation theory, but to date no works have computed
all third-order diagrams in neutron matter with three-body forces included. Moreover, single-particle 
propagators are often treated at the (first-order) Hartree-Fock level, which introduces a strong momentum dependence
in the mean-field potential that reduces the second- and third-order 
diagrammatic contributions to the energy per particle by up to 30\% and 50\%, respectively, at nuclear
matter saturation density. However, it is well known 
\cite{bertsch68,hebeler10,holt11npa} that second-order perturbative contributions to the nucleon
self-energy in nuclear matter from two-body forces
\cite{holt12} reduce the momentum dependence of the single-particle potential in the vicinity of the 
Fermi surface. A conservative estimate of this uncertainty can therefore be obtained by employing both a 
free-particle spectrum and a Hartree-Fock spectrum \cite{drischler16}. An additional aim of the present
work is to reduce this significant source of uncertainty.

We employ as a starting point realistic $N\!N$-potentials at different orders 
\{$(q/\Lambda_\chi)^2, (q/\Lambda_\chi)^3, (q/\Lambda_\chi)^4$\} in the chiral expansion, corresponding to
next-to-leading order (NLO), next-to-next-to-leading order (N2LO), and next-to-next-to-next-to-leading 
order (N3LO). The short-range contact terms are fitted to elastic nucleon-nucleon scattering phase 
shifts and deuteron properties, while the intermediate- and long-range interaction is determined 
uniquely by one- and multi-pion exchange processes constrained by chiral symmetry. We vary the 
momentum-space cutoff $\Lambda$, which sets the scale at which nuclear forces are resolved, 
over the range $\Lambda \simeq (400-500)$\,MeV \cite{entem03,coraggio07,marji13} suitable for many-body
perturbation theory calculations of the energy density. 
We include as well the chiral 
three-nucleon force at order $(q/\Lambda_\chi)^3$, which is fitted to the binding energy and beta-decay 
lifetime of $^3H$ \cite{sammarruca15}. 
Extending to a consistent treatment at order $(q/\Lambda_\chi)^4$ requires the refitting of the
$c_D$ and $c_E$ low-energy constants, which is currently a work in progress. In the following we
refer to this partly incomplete treatment at order $(q/\Lambda_\chi)^4$ as N3LO*. The coarse-resolution 
chiral potentials employed in the present work have also been used to study the response functions
of neutron matter \cite{davesneholt15} as well as numerous thermodynamic properties of symmetric
nuclear matter and neutron matter \cite{wellenhofer16} (for recent reviews see Refs.\ \cite{holt13ppnp,holt16pr}).

\begin{figure}[t]
\centerline{\includegraphics[width=6.5cm]{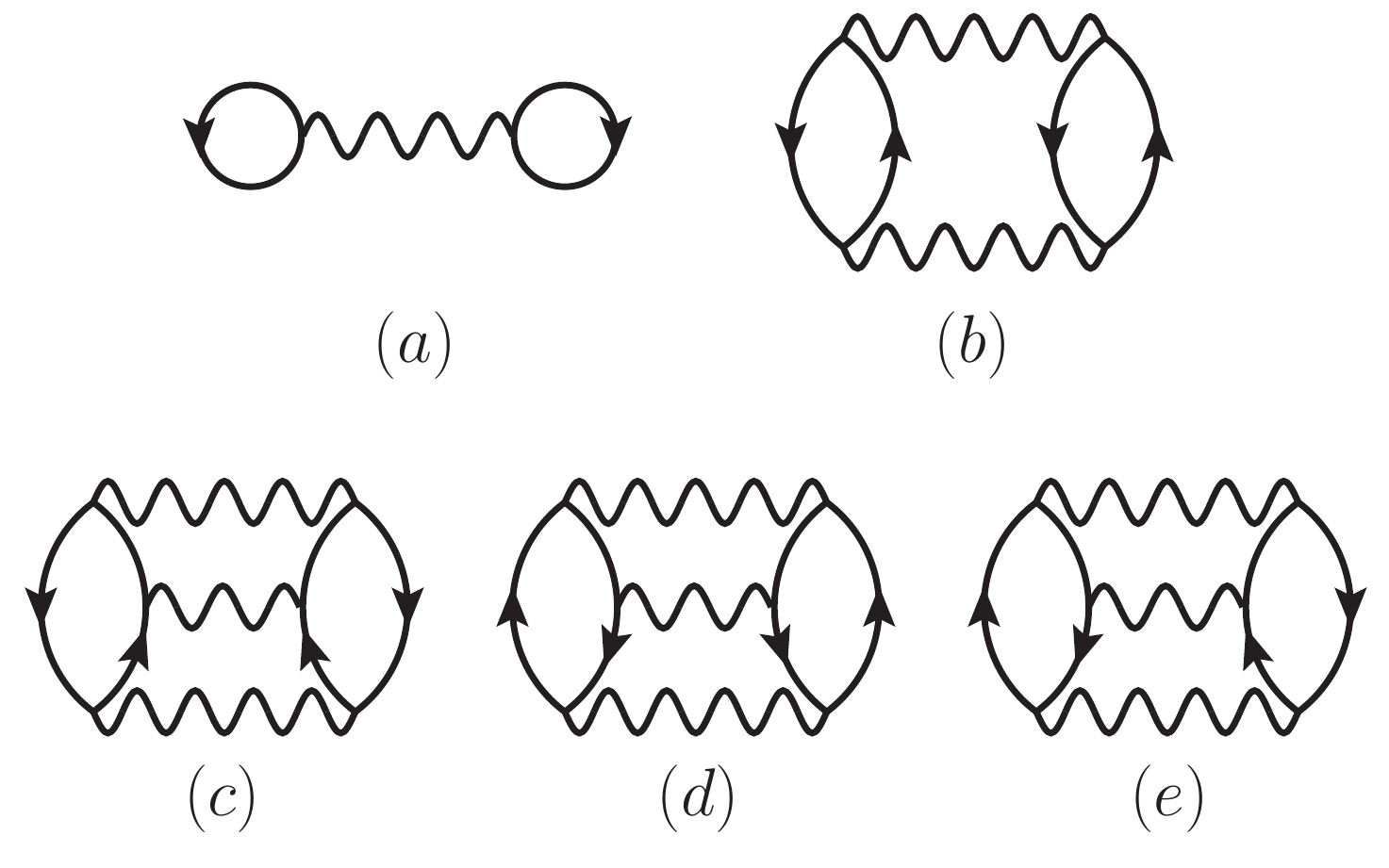}}
\caption{First-, second-, and third-order diagrammatic contributions to the ground-state energy density of isospin-symmetric nuclear matter and pure neutron 
matter from the (effective) chiral two-nucleon potential described in the text. 
The wavy line includes
the (antisymmetrized) density-dependent $N\!N$-interaction derived from the chiral three-body force at N2LO.} 
\label{gold}
\end{figure}

The present paper is organized as follows. In Section \ref{e3} we outline the calculation of the ground state
energy density of symmetric nuclear matter and pure neutron matter at third order in perturbation theory,
including self-consistent single-particle spectra. As a benchmark for the complicated partial-wave
decomposition of the third-order particle-hole ring-diagram for realistic $N\!N$-potentials, we 
present semi-analytic results for the direct terms from central and tensor 
model-type interactions. In 
Section \ref{results} we present a comprehensive study of the theoretical uncertainties for the equation of 
state of symmetric nuclear matter and pure neutron matter. We extract the isospin-asymmetry
energy $S_0$ and its slope parameter $L$ at saturation density, including also the results from
nuclear forces at NLO and N2LO in the chiral power counting. Furthermore, the 
third-order particle-hole diagrams from an $S$-wave contact-interaction 
(with two parameters $a_s$ and  $a_t$) 
allow us to examine the commonly used quadratic approximation 
in the isospin-asymmetry. We end with a summary and conclusions.

\section{Energy per particle at third order in perturbation theory}
\label{e3}

The first-, second-, and third-order perturbative contributions to the energy density of nuclear (or neutron)
matter are shown diagrammatically in Fig.\,\ref{gold}. The wavy line represents the (antisymmetrized)
density-dependent $N\!N$-interaction as given by the sum of the free-space $N\!N$-potential and the 
in-medium $N\!N$-interaction derived from the next-to-next-to-leading order (N2LO) chiral three-nucleon 
force. The latter is obtained by closing two external legs and summing over the filled Fermi sea of 
(free) nucleons \cite{holt09,hebeler10}. The method is equivalent to constructing a normal-ordered Hamiltonian
with respect to the noninteracting ground state and neglecting the residual three-body contribution
\cite{hagen07}.
This approximation has been improved in other works by including three-body forces at N3LO 
\cite{tews13} and by keeping the residual three-body force after normal ordering 
\cite{kaiser12,drischler16}. The first-, second-, and third-order contributions to the energy density 
$\rho E$ (with $\rho$ the density and $E$ the energy per particle) are given by

\begin{equation}
\rho E^{(1)} = \frac{1}{2}\sum_{12} \,
n_1 n_2 \langle 1 2 \left | (\overline{V}_{NN}+ \overline{V}_{NN}^{\text{med}}/3)\right | 1 2 \rangle,
\label{e1}
\end{equation}

\begin{equation}
\rho E^{(2)} = -\frac{1}{4} 
\sum_{1234} \left| \langle 1 2 \left | \overline{V}_{\text{eff}} \right | 3 4 \rangle \right |^2
\frac{n_1 n_2 \bar{n}_3 \bar{n}_4}
{e_3+e_4-e_1-e_2},
\label{e2}
\end{equation}

\begin{eqnarray}
\rho E^{(3)}_{pp} &=& \frac{1}{8} 
\sum_{123456} \langle 1 2 \left | \overline{V}_{\text{eff}} \right | 3 4 \rangle
\langle 3 4 \left | \overline{V}_{\text{eff}} \right | 5 6 \rangle
\langle 5 6 \left | \overline{V}_{\text{eff}} \right |  1 2 \rangle \nonumber \\ 
&& \times \frac{n_1 n_2 \bar{n}_3 \bar{n}_4 \bar{n}_5 \bar{n}_6}
{(e_3+e_4-e_1-e_2)(e_5+e_6-e_1-e_2)},
\label{e3pp}
\end{eqnarray}

\begin{eqnarray}
\rho E^{(3)}_{hh} &=& \frac{1}{8} 
\sum_{123456} \langle 1 2 \left | \overline{V}_{\text{eff}} \right | 3 4 \rangle
\langle 3 4 \left | \overline{V}_{\text{eff}} \right | 5 6 \rangle
\langle 5 6 \left | \overline{V}_{\text{eff}} \right |  1 2 \rangle \nonumber \\ 
&& \times \frac{\bar{n}_1 \bar{n}_2 n_3 n_4 n_5 n_6}
{(e_1+e_2-e_3-e_4)(e_1+e_2-e_5-e_6)},
\label{e3hh}
\end{eqnarray}

\begin{eqnarray}
\rho E^{(3)}_{ph} &=& 
-\sum_{123456} \langle 1 2 \left | \overline{V}_{\text{eff}} \right | 3 4 \rangle
\langle 5 4 \left | \overline{V}_{\text{eff}} \right | 1 6 \rangle
\langle 3 6 \left | \overline{V}_{\text{eff}} \right | 5 2 \rangle \nonumber \\ 
&& \times \frac{n_1 n_2 \bar{n}_3 \bar{n}_4 n_5 \bar{n}_6}
{(e_3+e_4-e_1-e_2)(e_3+e_6-e_2-e_5)},
\label{e3ph}
\end{eqnarray}

\noindent where $n_j=\theta(k_f-|\vec p_j|)$ is the (step-like) distribution function, $\bar n_j=1-n_j$, 
$\overline{V}= V-{\cal P}_{12}V$ is the (Fierz) antisymmetrized $N\!N$-potential
with ${\cal P}_{12}$ the exchange-operator in spin-, isospin- and 
momentum-space. The effective $N\!N$-potential is given by the sum 
$V_{\text{eff}} = V_{NN}+ V_{NN}^{\text{med}}$. Note that the third-order particle-particle
and hole-hole Goldstone diagrams have a symmetry factor of $\frac{1}{8} = \frac{1}{2^3}$
arising from three pairs of equivalent lines, while the third-order particle-hole diagram has no
equivalent pairs of lines and consequently
an overall symmetry factor of 1.

\begin{figure}[t]
\centerline{\includegraphics[width=7cm]{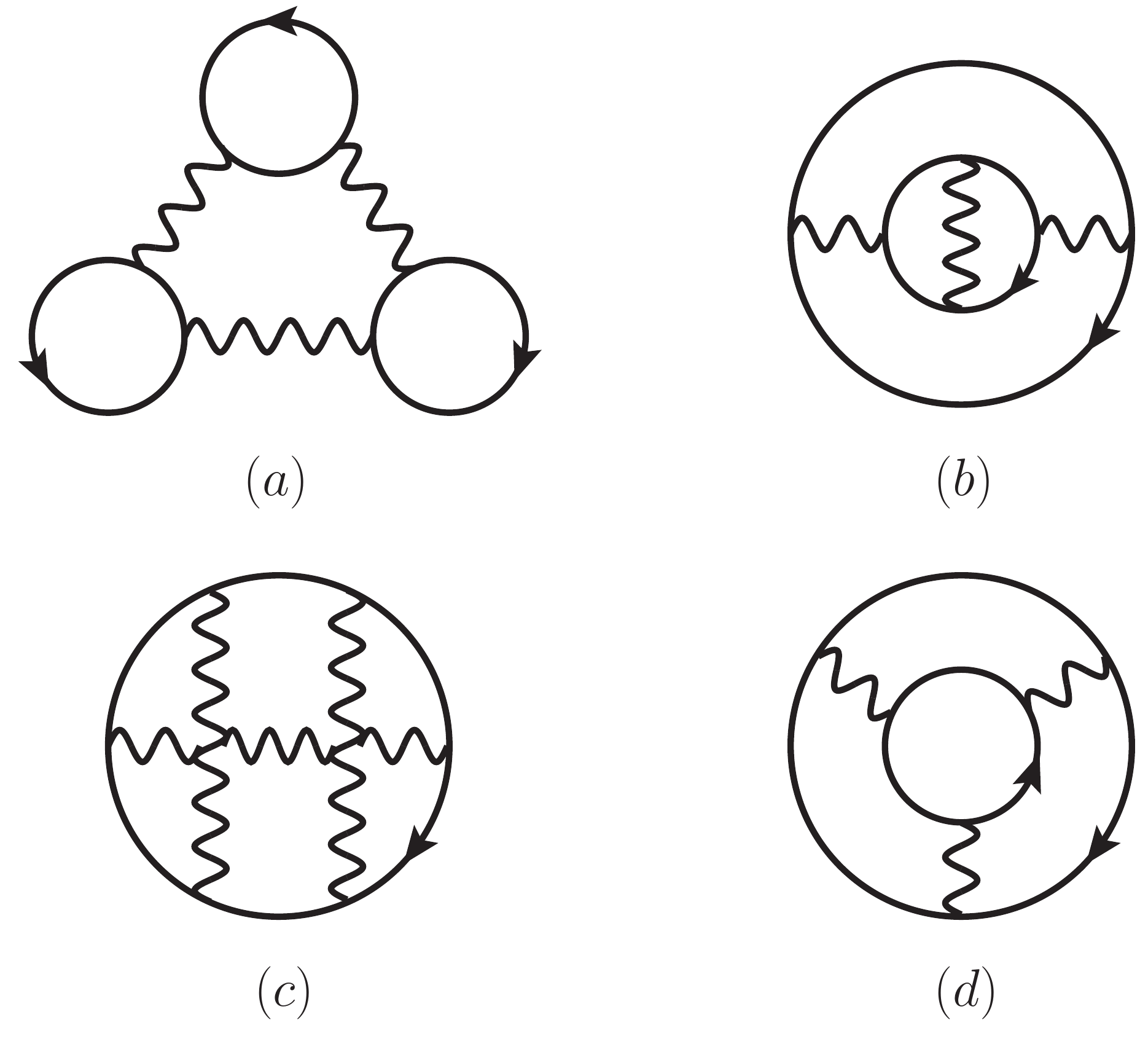}}
\caption{Four ring-diagrams representing the third-order particle-hole 
contribution, organized according to the number of direct and exchange 
interactions.  Diagrams (a), (b), (c) and (d) have 0, 1, 2, and 3 exchange 
interactions, respectively.} 
\label{e3phfig}
\end{figure}

The third-order particle-particle 
and hole-hole contributions can be straightforwardly decomposed in terms of partial-wave matrix elements of 
$\overline V$ and written 
as integrals over the relative momenta of the interacting nucleons. The third-order particle-hole
contribution, on the other hand, is more conveniently calculated by integrating over the 
individual particle-momenta, which however leads to more complicated expressions when 
written in terms of partial-wave matrix elements. We therefore provide semi-analytical
expressions for several of the third-order particle-hole ring diagrams from model-type interactions, which 
are useful to benchmark the results of extensive numerical calculations. We begin by decomposing the 
third-order particle-hole ring contribution into four parts, shown in Fig.\ \ref{e3phfig},
according to the number of direct $(dir)$ and exchange $(exch)$ interactions. We denote diagram 
(a) in Fig.\ \ref{e3phfig} as the $dir^3$ term, diagram (b) as the $dir^2 \cdot exch$ term,
diagram (c) as the $dir \cdot exch^2$ term, and diagram (d) as the $exch^3$ term. We have
computed all contributions (a)$-$(d) for various model-type interactions and one-pion exchange, but
the parts involving multiple exchange terms lead to very lengthy expressions, and for brevity we 
present the semi-analytic results here only for diagrams (a) and (b) in Fig.\ \ref{e3phfig}.

We consider first a scalar isoscalar boson-exchange interaction of the form
\begin{equation}
V_{\rm dir}(q) = -{g^2  \over m^2+q^2}\,,
\label{scal}
\end{equation}
with $q$ the momentum transfer. The contribution to the energy per particle $E(\rho)$ of symmetric nuclear 
matter from diagram $(a)=dir^3/6$ is given by
\begin{equation}
E(\rho)^{\rm (a)} = - {g^6 M^2 \over 32 \pi^7 k_f} 
\int\limits_0^\infty\!ds \!\int\limits_0^\infty\!d\kappa \bigg[{Q_0(s,\kappa) 
\over s^2+\beta}\bigg]^3\,,
\end{equation}
where $M$ is the nucleon mass, $\beta = m^2/4k_f^2$, and the Fermi momentum is related to the density by 
$\rho = 2k_f^3/3\pi^2$. The (Euclidean) polarization function $Q_0(s,\kappa)$, 
arising from an individual nucleon-ring in diagram $(a)$, has the following analytical form
\begin{eqnarray}
Q_0(s,\kappa) &=& s - s \kappa \arctan \frac{1+s}{\kappa} - s\kappa \arctan \frac{1-s}{\kappa} \nonumber \\
&+& \frac{1}{4}(1-s^2+\kappa^2) \ln{(1+s)^2+
\kappa^2 \over (1-s)^2+\kappa^2}\,.
\end{eqnarray}
The contribution to the energy per particle of symmetric nuclear matter from diagram 
$(b)=-dir^2\cdot exch/2$ can be represented  by a six-fold integral
\begin{eqnarray}
E(\rho)^{\rm (b)} &=&  {6g^6 M^2 \over (2\pi)^7 k_f}
\int\limits_0^\infty\!\!ds \!\!\int\limits_0^\infty\!\!d\kappa\!\!\int
\limits_0^1\!\!dl_1 \!\!\int\limits_0^1\!\!dl_2\!\!\int\limits_{-l_1}^{l_1}\!\!dx 
\!\!\int\limits_{-l_2}^{l_2}\!\!dy \, \nonumber \\
&& \hspace{-.7in} {l_1l_2\,Q_0(s,\kappa)\,(s^2+\beta)^{-2}
\over [(s+x)^2+\kappa^2][(s+ y)^2+\kappa^2]} \Big\{
\big[(s+ x)(s+y)-\kappa^2\big] \nonumber \\
&& \hspace{-.7in} \times W_a^{-1/2} +\big[(s+ x)(s+y)+\kappa^2\big]W_b^{-1/2} \Big\}\,, 
\end{eqnarray}
with the auxiliary functions $W_a= \big[4\beta+l_1^2+l_2^2-2x y\big]^2-4(l_1^2-x^2)(l_2^2-y^2)$ 
and $W_b=\big[4\beta+l_1^2+l_2^2+4s(s+x+y)+2x y\big]^2-4(l_1^2-x^2)(l_2^2-y^2)$.

For a modified pseudoscalar isovector boson-exchange interaction of the form
\begin{equation}
V_{\rm dir} = -g^2 \vec \tau_1\cdot\vec\tau_2 \, {\vec \sigma_1\cdot\vec q
\,\vec \sigma_2\cdot\vec q \over (m^2+q^2)^2},
\label{mpe}
\end{equation}
the contribution to the energy per particle of symmetric nuclear matter from diagram 
$(a)=dir^3/6$ reads
\begin{equation}
E(\rho)^{\rm (a)} = - {3g^6 M^2 \over 32 \pi^7 k_f} 
\int\limits_0^\infty\!ds \!\int\limits_0^\infty\!d\kappa \bigg[{s^2 Q_0(s,\kappa) 
\over (s^2+\beta)^2}\bigg]^3\,.
\end{equation}
The contribution to the energy per particle of symmetric nuclear matter from diagram 
$(b)=-dir^2\cdot exch/2$ is on the other hand given by
\begin{eqnarray}
E(\rho)^{\rm (b)} &=&  {18g^6 M^2 \over (2\pi)^7 k_f}
\int\limits_0^\infty\!\!ds \!\!\int\limits_0^\infty\!\!d\kappa\!\!\int
\limits_0^1\!\!dl_1 \!\!\int\limits_0^1\!\!dl_2\!\!\int\limits_{-l_1}^{l_1}\!\!dx 
\!\!\int\limits_{-l_2}^{l_2}\!\!dy \, \nonumber \\
&&\hspace{-.7in}{l_1l_2s^4\,Q_0(s,\kappa)\,(s^2+\beta)^{-4}
\over [(s+x)^2+\kappa^2][(s+ y)^2+\kappa^2]}\bigg\{\big[(s+ x)(s+y)-\kappa^2\big]\nonumber \\ 
&& \hspace{-.8in} \times \Big[4\beta(l_1^2+l_2^2-2x^2-2y^2+2xy) 
+(l_1^2-l_2^2-2x^2+2xy) \nonumber \\
&&\hspace{-.8in}\times(l_1^2-l_2^2+2y^2-2xy)\Big] W_a^{-3/2}
+\big[(s+ x)(s+y)+\kappa^2\big] \nonumber  \\
&&\hspace{-.8in}\times \Big[4\beta\big(l_1^2+l_2^2-4s(s+x+y)-2x^2-2y^2-2xy\big) \nonumber \\ 
&& \hspace{-.8in} +\big(l_1^2-l_2^2-4s(s+x+y)-2x(x+y)\big) \nonumber \\
&& \hspace{-.8in} \times \big(l_1^2-l_2^2+4s(s+x+y)+2y(x+y)
\big)\Big] W_b^{-3/2} \bigg\}\,.
\end{eqnarray}

\begin{figure}[t]
\centerline{\includegraphics[width=6cm,angle=270]{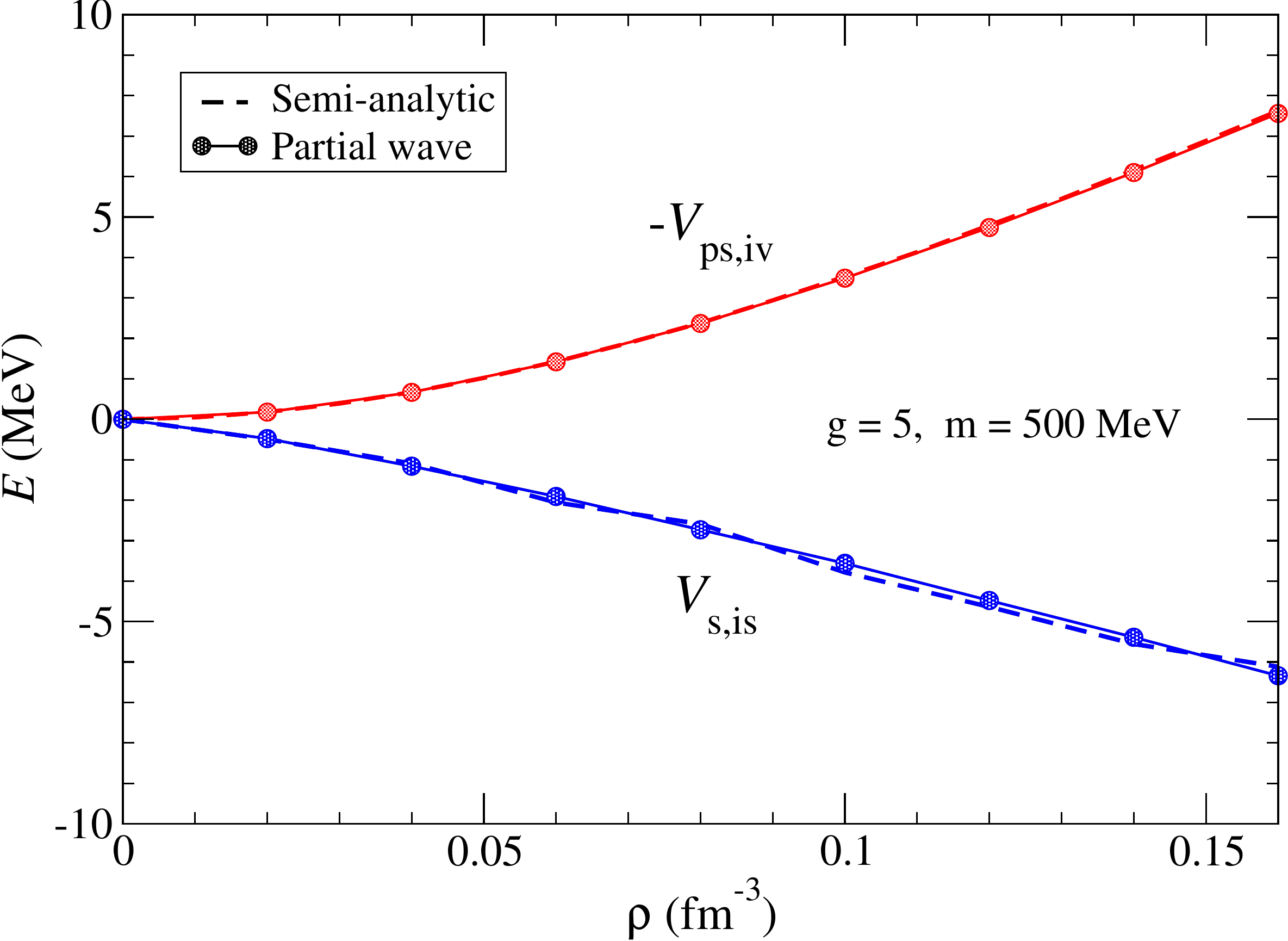}}
\caption{(Color online) The density dependence of the third-order particle-hole contribution to the 
ground state energy per particle, $E$, of symmetric nuclear matter from the 
scalar-isoscalar test interaction in Eq.\ (\ref{scal}) and the modified 
pseudoscalar-isovector test interaction in Eq.\ (\ref{mpe}). The contribution from the latter 
is multiplied by -1 for better clarity in the figure. The results of
numerical calculations based on a partial-wave decomposition are shown together with
semi-analytic results involving only multidimensional integrals.} 
\label{test}
\end{figure}

In Fig.\ \ref{test} we compare the sum of all four contributions $(a)\!+\!(b)\!+\!(c)\!+\!(d)$ to the 
third-order particle-hole diagram in symmetric nuclear matter from the two test
interactions in Eqs.\ (\ref{scal}) and (\ref{mpe}) employing both a partial-wave 
decomposition (used in Section \ref{results} below for realistic chiral two- and three-nucleon 
forces) as well as a semi-analytic evaluation. The boson mass is taken to be
$m = 500$\,MeV for both interactions and the coupling constant is chosen to be $g=5$.
For the modified pseudoscalar-isovector force model, labeled $V_{ps,iv}$ in Fig.\ \ref{test},
we have changed the overall sign of the interaction to more easily differentiate the two terms
in the figure. We see from Fig.\ \ref{test} that the two sets 
of calculations agree almost perfectly. In the case of the scalar-isoscalar test interaction, 
large cancellations among the four terms computed with the semi-analytic 
method result in numerical noise 
on the order of (4-8)\% relative to the results from the partial-wave decomposition. In comparison, 
the two methods agree to within 1\% across all densities for the energy per particle of 
symmetric nuclear matter from the modified isovector-pseudoscalar test interaction. 
A further case to check our numerical calculations is given by the third-order ring contributions from an 
$S$-wave contact-interaction of the form $V_{\text{ct}} = -{\pi \over M}[a_s+3a_t+(a_t-a_s)\vec \sigma_1\!\cdot
\!\vec \sigma_2]$. Using the polarization function $Q_0(s,\kappa)$ in Eq.\ (8) and integrating over its cube, 
one finds for symmetric nuclear matter
\begin{equation}  E(\rho) = 1.04814\,{k_f^5 \over \pi^4 M}(a_s+a_t)(5a_s^2-14a_sa_t+5a_t^2)\,,\end{equation}  
and for pure neutron matter
\begin{equation}  E_n(\rho_n) = 2.79505\,{k_n^5\over \pi^4 M}\, a_s^3\,,\end{equation}  
with $k_n$ the neutron Fermi momentum related to the neutron density by $\rho_n = k_n^3/3\pi^2$. For various 
choices of the scattering lengths $a_s$ and $a_t$, we reproduce these results with an accuracy of 
$1\%$ and better.

Until now we have assumed a free-particle spectrum for the single-particle energies occurring 
in the denominators of the second- and third-order perturbative contributions to the equation of 
state. For calculations involving high-precision chiral two- and three-body forces, it is 
convenient to employ Hartree-Fock single-particle energies, which are well approximated by 
the effective mass $M^*$ plus energy shift $\Delta$ parametrization
\begin{equation}
e(p) = p^2/2M^* + \Delta,
\label{hfspe}
\end{equation}
where $\Delta$ is independent of the momentum $p$. At the mean field level, 
the single-particle potential has a strong momentum dependence that gives rise to nucleon 
effective masses in the vicinity of $M^*/M \simeq 0.7$ at nuclear matter saturation density.
This leads to a reduction of the second-order energy contribution in Eq.\ (\ref{e2}) by roughly
30\%. All third-order contributions are likewise scaled by $(M^*/M)^2 \simeq 0.5$ at nuclear
saturation density. A second-order perturbative treatment of the nucleon self-energy 
in symmetric nuclear matter, however, gives rise to an effective mass that is itself
strongly momentum dependent, and the parametrization in Eq.\ (\ref{hfspe}) is no longer valid.
In particular, close to the Fermi momentum the effective mass peaks at a value close to
the free-space mass \cite{bertsch68}. The associated uncertainty in the symmetric nuclear
matter equation of state at saturation density is on the order of 5\,MeV \cite{coraggio14} 
while for neutron matter the uncertainty is 2-3\,MeV \cite{coraggio13}.

\begin{figure}[t]
\centerline{\includegraphics[width=5.7cm,angle=270]{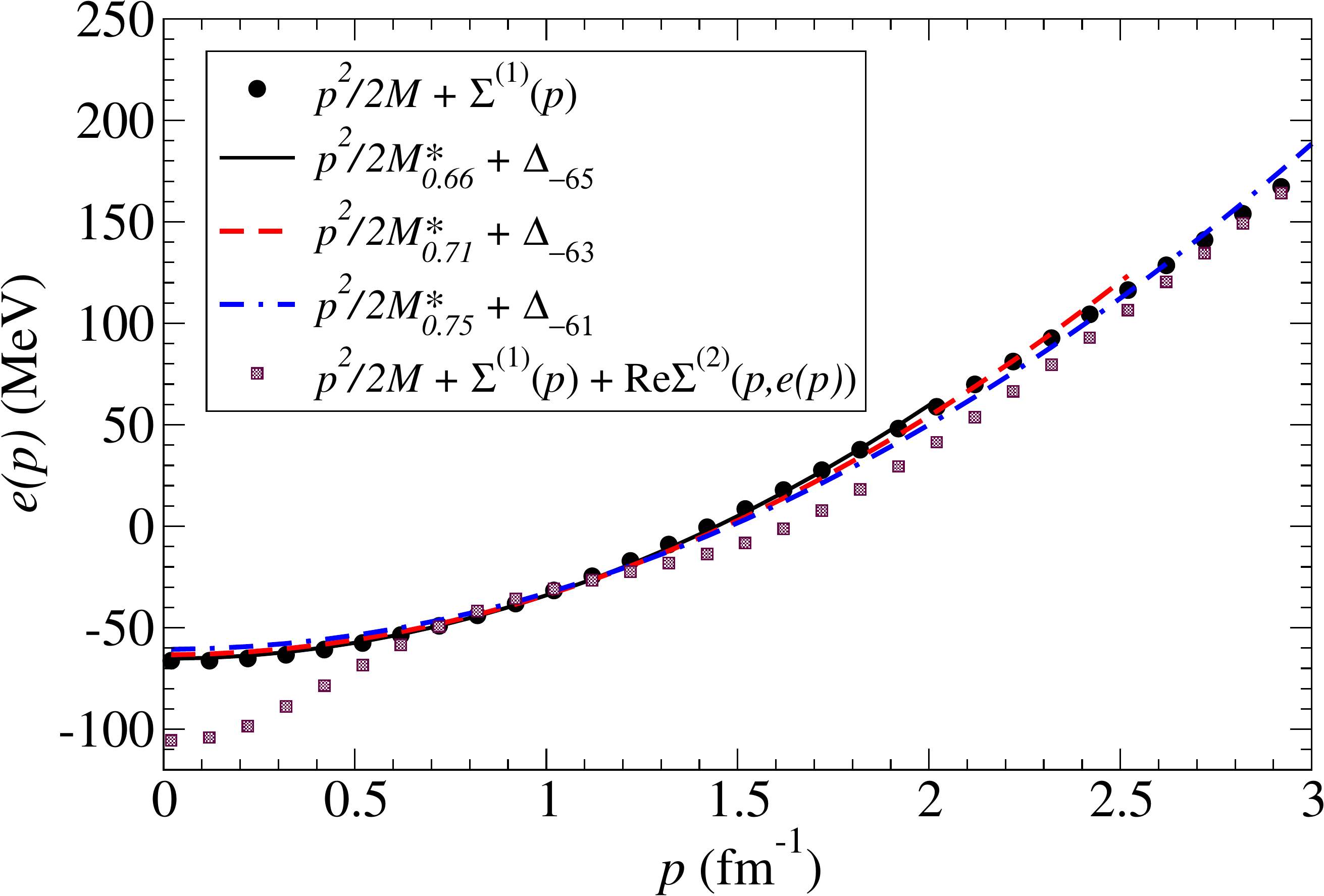}}
\caption{Nucleon single-particle energy in symmetric nuclear matter including the 
first-order $\Sigma^{(1)}$ and second-order $\Sigma^{(2)}$ perturbative contributions 
to the self-energy. The effective mass and energy shift parametrization is shown for
three different choices of the fitting region in momentum space, and the values of the
effective mass ($M^*/M$) and energy shift $\Delta$ (in units of MeV) are shown as subscripts.} 
\label{spe1}
\end{figure}

In Fig.\ \ref{spe1} we show the single-particle energy calculated at the Hartree-Fock level,
$e(p) = p^2/2M + \Sigma^{(1)}(p)$, together with three different effective mass plus
energy shift parametrizations fitted over the ranges in momentum: $p < 2.0$\,fm$^{-1}$,
$p < 2.5$\,fm$^{-1}$, and $p < 3.0$\,fm$^{-1}$. The subscripts on the $M^*$ and $\Delta$
terms refer to the values of ($M^*/M$) and $\Delta$ (in units of MeV). The momentum 
dependence of the single-particle energy at the Hartree-Fock level remains nearly quadratic,
but we observe that the values of $M^*/M$ and $\Delta$ depend on the choice of the
fitting region.

In Section \ref{results} below, we compute the single-particle energies
in Eqs.\ $(\ref{e2})-(\ref{e3ph})$ self-consistently at second order:
\begin{equation}
e(p) = \frac{p^2}{2M} + \Sigma^{(1)}(p) + {\rm Re}\,\Sigma^{(2)}(p,e(p)).
\label{scspe}
\end{equation}
The Hartree-Fock contribution, $\Sigma^{(1)}(p)$,
to the nucleon self-energy in nuclear matter depends only on the momentum and is manifestly real.
The second-order contribution, $\Sigma^{(2)}(p,e(p))$, is in general complex and energy dependent.
In practice Eq.\ (\ref{scspe}) is solved iteratively until a converged solution is reached. The inclusion
of a density-dependent two-body force derived from the leading chiral three-body force requires
an additional symmetry factor of 1/2 in the Hartree-Fock contribution $\Sigma^{(1)}(p)$.
For additional computational details we refer the reader to Refs.\ \cite{holt13prc,holt16prc}. 
We show in Fig.\ \ref{spe1} the nucleon single-particle energy in symmetric nuclear matter including
also the second-order contribution $\Sigma^{(2)}(p,e(p))$ to the self energy. In contrast to the 
Hartree-Fock approximation, the momentum dependence of the single-particle energy is no
longer approximately quadratic.

The nucleon self energy in infinite nuclear matter is related to the volume 
components of the nucleon-nucleus optical potential probed in elastic scattering experiments. 
If Refs.\ \cite{holt13prc,holt16prc} we
employed low-momentum chiral two- and three-body forces at second-order in perturbation 
theory and found very good agreement with the isoscalar and isovector real components of the 
optical potential compared to phenomenology \cite{becchetti69,varner91,koning03} for energies
$E\lesssim 200$\,MeV.

\section{Results}
\label{results}

\begin{figure}[t]
\centerline{\includegraphics[width=6cm,angle=270]{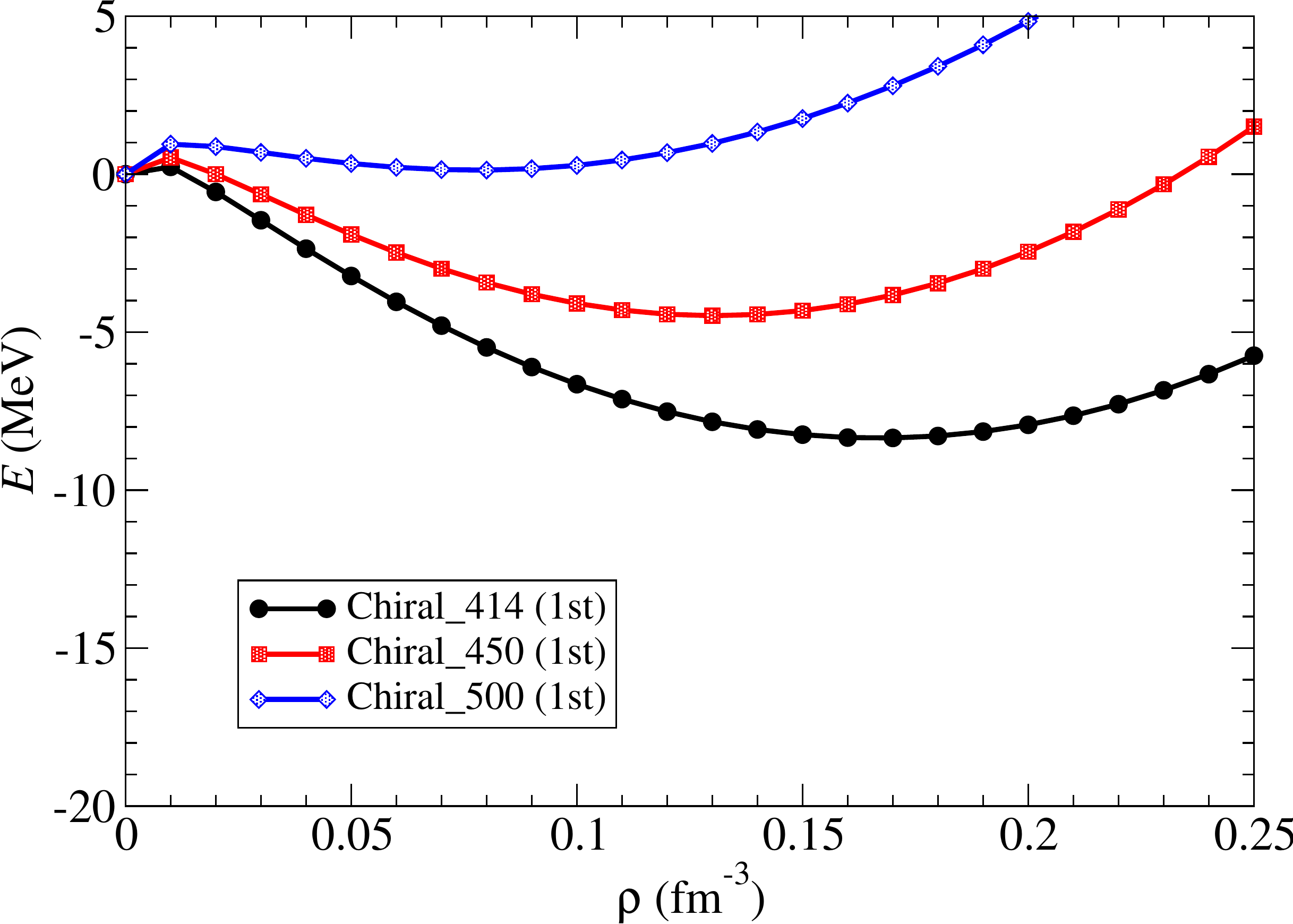}}
\caption{(Color online) First-order diagrammatic contribution to the energy per particle of symmetric 
nuclear matter from N3LO* chiral two- and three-nucleon forces defined at different resolution scales:
$\Lambda =$ 414, 450, and 500\,MeV.} 
\label{e_1_iso0}
\end{figure}

In this section we calculate the energy per particle of symmetric nuclear matter and
pure neutron matter at third order in perturbation theory with self-consistent single-particle energies
at second order. Our aim is to provide improved theoretical
error estimates on the equations of state $E(\rho)$ and $E_n(\rho_n)$, the density-dependent symmetry 
energy $S(\rho)$, and the slope parameter $L$ of the symmetry energy. We account for theoretical uncertainties 
arising from the  convergence in perturbation theory, the choice of resolution scale, and the omission of 
higher-order terms in the chiral expansion.

We begin with the equation of state of symmetric nuclear matter at first order in perturbation theory, 
shown in Fig.\ \ref{e_1_iso0} as a function of density for three different chiral potentials with 
momentum-space cutoffs of $\Lambda =$ 414, 450, and 500\,MeV. Each of the N3LO two-nucleon 
forces are supplemented with a density-dependent $N\!N$-interaction constructed from the N2LO 
three-body force, whose low-energy constants $c_D$ and $c_E$ are fitted to the binding energy 
and lifetime of the triton. In Table \ref{snmtab} we also 
show the specific values of the first-order contribution (in units of MeV) at nuclear matter saturation 
density $\rho_0$. For comparison the noninteracting Fermi gas contribution (i.e.
the kinetic energy 
$E_{\text{kin}}(\rho)=3k_f^2/10M$) at this density is $E_{\text{kin}} = 22.1$\,MeV.  At the mean-field 
level there is a large uncertainty associated with the choice of resolution scale, a feature observed
already in Ref.\ \cite{bogner05}. However, the error band at this order in the perturbative expansion
does not pass through the empirical saturation point at $\rho_0 = 0.16$\,fm$^{-3}$ and 
$E(\rho_0) = -16$\,MeV, and therefore varying the resolution scale over the range typically 
chosen in constructing chiral potentials does not encompass the full theoretical uncertainty. 

\begin{figure}[t]
\centerline{\includegraphics[width=6cm,angle=270]{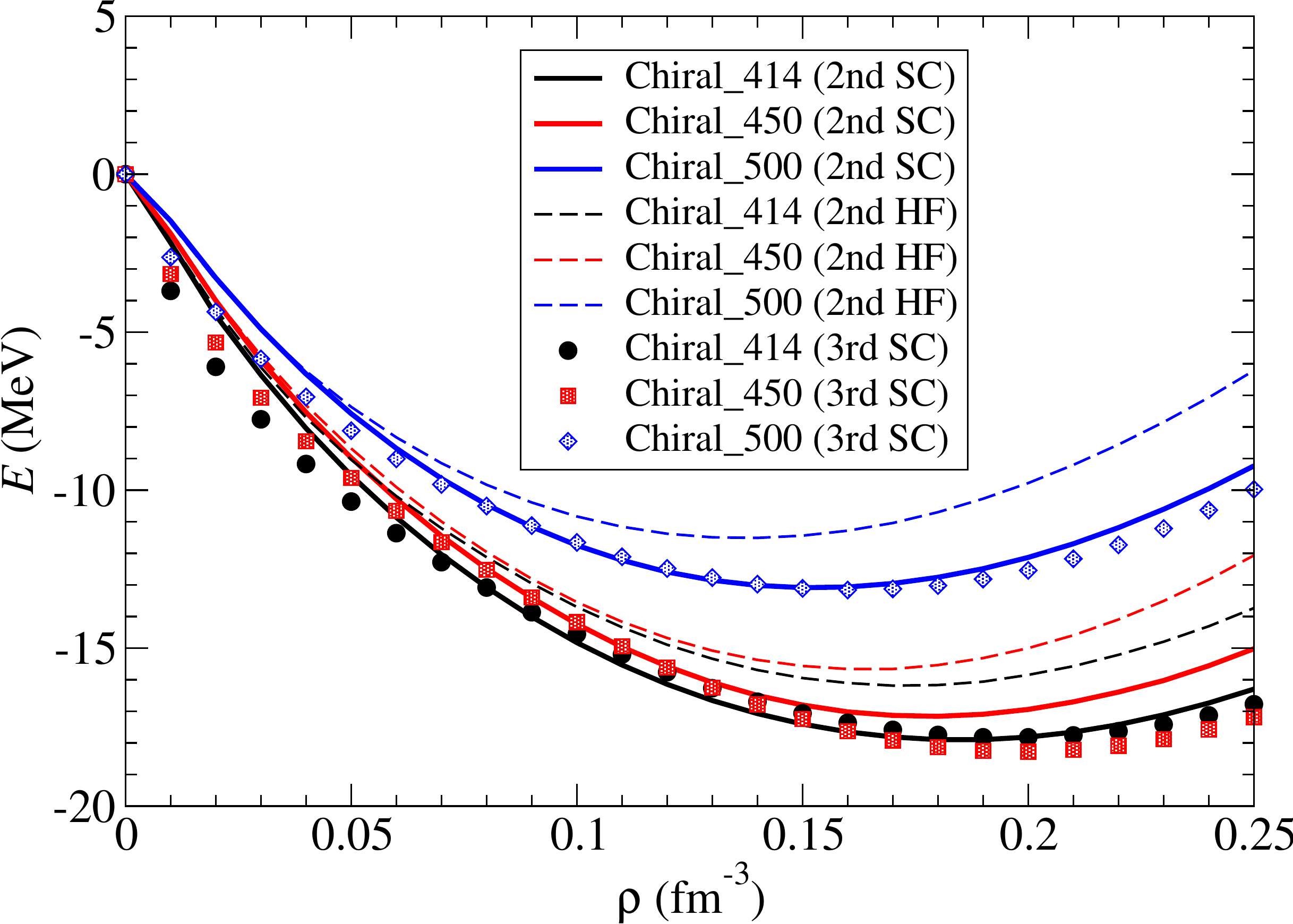}}
\caption{(Color online) The equation of state of symmetric nuclear matter at second and third order in 
many-body perturbation theory from N3LO* chiral two- and three-body forces. Both a Hartree-Fock (HF) and
self-consistent (SC) second-order single-particle spectrum were employed for the three chiral 
potentials with different momentum-space regulator scales.} 
\label{e_1_2_3_iso0}
\end{figure}

The results for the equation of state of symmetric nuclear matter including second-order
perturbative contributions are shown in Fig.\ 
\ref{e_1_2_3_iso0}. We consider both a Hartree-Fock spectrum (labeled ``2nd HF'') for the 
intermediate-state energies as well as a self-consistent second-order approximation (labeled
``2nd SC''). As expected, the latter leads to additional attraction resulting from a reduced 
momentum-dependence of the single-particle potential. The differences between the 
second-order contributions with a HF and SC spectrum reach up to (2-3)\,MeV for 
densities below $\rho=0.25$\,fm$^{-3}$, a feature that is largely independent of 
the choice of resolution scale. In Table \ref{snmtab} we show 
the specific values at nuclear matter saturation density and include also the results, labeled $E^{(2)}$, 
employing a free-particle spectrum.
By comparing the magnitude of the second-order contributions $E^{(2SC)}$ to the leading-order 
Hartree-Fock contributions $E^{(1)}$ at nuclear matter saturation density, we observe an 
improved convergence pattern as the momentum-space cutoff $\Lambda$ is lowered. At second order
in perturbation theory, the error estimate obtained through varying the cutoff scale now encompasses
the empirical saturation point.

The third-order contributions to the energy per particle of symmetric nuclear matter, with single-particle 
energies computed self consistently at second order, are
shown in Fig.\,\ref{e_1_2_3_iso0} and labeled ``3rd SC''. We observe that taken together the three
contributions $E^{(3)}_{pp}$, $E^{(3)}_{hh}$, and $E^{(3)}_{ph}$ give rise to 
additional attraction at both low and high densities. In particular, for
densities less than $\rho \simeq 0.08$\,fm$^{-3}$, where a spinodal instability is expected 
\cite{wellenhofer15}, the third-order terms cannot be neglected. 
At and above saturation density, the perturbation theory expansion appears to be better
converged, though generically both repulsive particle-particle and attractive particle-hole 
contributions are individually on the order of (1-3)\,MeV. In Table \ref{snmtab} 
we show the values of the three third-order contributions at nuclear matter saturation density
including self-consistent second-order single-particle energies, for each of the three chiral
potentials considered in this section. Given the systematic cancellations that occur between 
the third-order particle-particle and particle-hole diagrams 
(independent of resolution scale $\Lambda$), we suggest that these terms should be included 
together or not at all. From Fig.\,\ref{e_1_2_3_iso0} we see that the largest source of 
theoretical uncertainty comes from the choice of resolution scale, as was found previously in Ref.\ 
\cite{sammarruca15}. The empirical saturation point is nearly at the central value of the 
error band, but there remains a $\Delta E \simeq 6$\,MeV uncertainty in the energy
per particle at $\rho = \rho_0$.

\setlength{\tabcolsep}{.035in}
\begin{table}[t]
\begin{tabular}{|c|c|c|c|c|c|c|c|} \hline
\multicolumn{1}{|c|}{} & \multicolumn{1}{c|}{} & \multicolumn{3}{c|}{$E^{(2)}$}
& \multicolumn{3}{c|}{$E^{(3)}$} \\ \hline 
$\Lambda$ & $E^{(1)}$ & $E^{(2)}$ & $E^{(2HF)}$ & $E^{(2SC)}$ 
& $E^{(3SC)}_{pp}$ & $E^{(3SC)}_{hh}$ & $E^{(3SC)}_{ph}$ \\ \hline 
414 & $-30.1$ & $-11.0$ & $-7.9$   & $-9.7$   & $0.8$ & $-0.3$ & $-0.3$ \\ \hline
450 & $-25.9$ & $-15.9$ & $-11.5$ & $-13.2$ & $1.0$ & $-0.2$ & $-1.5$ \\ \hline
500 & $-19.5$ & $-18.7$ & $-13.3$ & $-15.7$ & $2.2$ & $-0.1$ & $-2.1$ \\ \hline
\end{tabular}
\caption{The contribution to the energy per particle $E(\rho_0)$ of symmetric nuclear matter from the first-,
second-, and third-order perturbation theory diagrams employing chiral two- and three-nucleon forces. 
For the second-order contribution we list the values using
a free single-particle spectrum, $E^{(2)}$, a Hartree-Fock spectrum, $E^{(2HF)}$, and 
self-consistent single-particle energies at second order, $E^{(2SC)}$. 
All values are in units of MeV and the noninteracting contribution (not included) is $E_{\text{kin}} = 
22.1$\,MeV.}
\label{snmtab}
\end{table}

\begin{figure}[t]
\centerline{\includegraphics[width=6cm,angle=270]{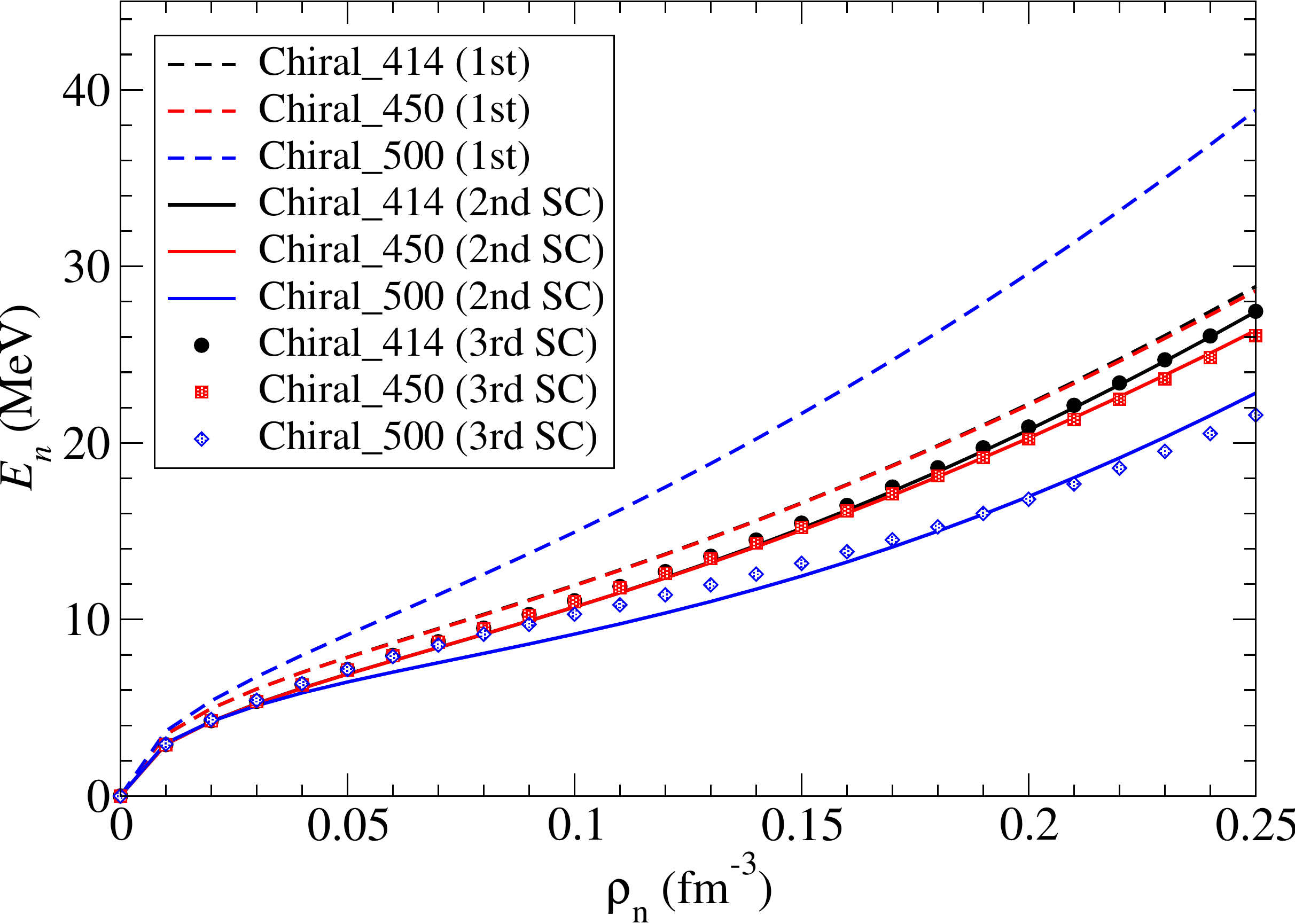}}
\caption{(Color online) Neutron matter equation of state at first, second, and third order in many-body
perturbation theory from N3LO* chiral two- and three-nucleon forces. For both the second- and
third-order contributions, the nucleon self-energies are computed self-consistently (SC) at 
second-order. Results are shown for chiral potentials constructed with a range of
momentum-space cutoffs: $\Lambda = (414, 450, 500)$\,MeV.} 
\label{e_mbpt_10}
\end{figure}

We next consider the equation of state of pure neutron matter from chiral two- and three-nucleon 
forces at order N3LO*. In Fig.\ \ref{e_mbpt_10} we 
show the results at first, second, and third order in perturbation theory from chiral potentials
with momentum-space cutoffs $\Lambda = (414,450,500)$\,MeV. At leading-order in perturbation
theory there is again a large dependence on the choice of resolution scale, but at both
second and third order, the variations are about $\Delta  E_n \simeq 2$\,MeV at 
$\rho_n = 0.16$\,fm$^{-3}$ and $\Delta  E_n \simeq 3$\,MeV at 
$\rho_n = 0.25$\,fm$^{-3}$. The inclusion of the third-order diagrams has relatively little effect
for the chiral potentials with $\Lambda = 414$\,MeV and $450$\,MeV, which we see give nearly
identical equations of state at each order in perturbation theory across all densities 
considered. In contrast, the equation
of state from the $\Lambda =500$\,MeV potential receives important contributions at low
densities that significantly reduces the scale dependence. In fact,
for densities up to $\rho_n \simeq 0.10$\,fm$^{-3}$, all N3LO* chiral potentials give a nearly unique
neutron matter equation of state. At higher densities the third-order contributions do
not reduce the scale dependence in any meaningful way.

Theoretical uncertainties on the neutron matter equation of state estimated from the convergence
pattern of many-body perturbation theory and variations in the resolution scale are relatively small. 
The error estimates on the density-dependent isospin-asymmetry energy, defined as
the difference $S(\rho) = E_n(\rho) - E(\rho)$, are
correspondingly tight, similar to what has already been reported in previous studies with microscopic
two- and three-body forces \cite{hebeler10prl,gandolfi12}, which predict values of the 
isospin-asymmetry energy and slope parameter that lie just outside of 
the experimental uncertainty band \cite{lattimer13,lattimer14}. To better understand this discrepancy
we consider now the errors due to neglected higher-order contributions in the chiral 
expansion. In particular, three- and four-body forces at N3LO are neglected in the present treatment
as well as N4LO two- and many-body forces. The two-body forces at N4LO have been shown to improve 
significantly in particular the $N\!N$-scattering phase shifts in $F$ and $G$ partial waves \cite{entem15}.

\begin{figure}[t]
\centerline{\includegraphics[width=6cm,angle=270]{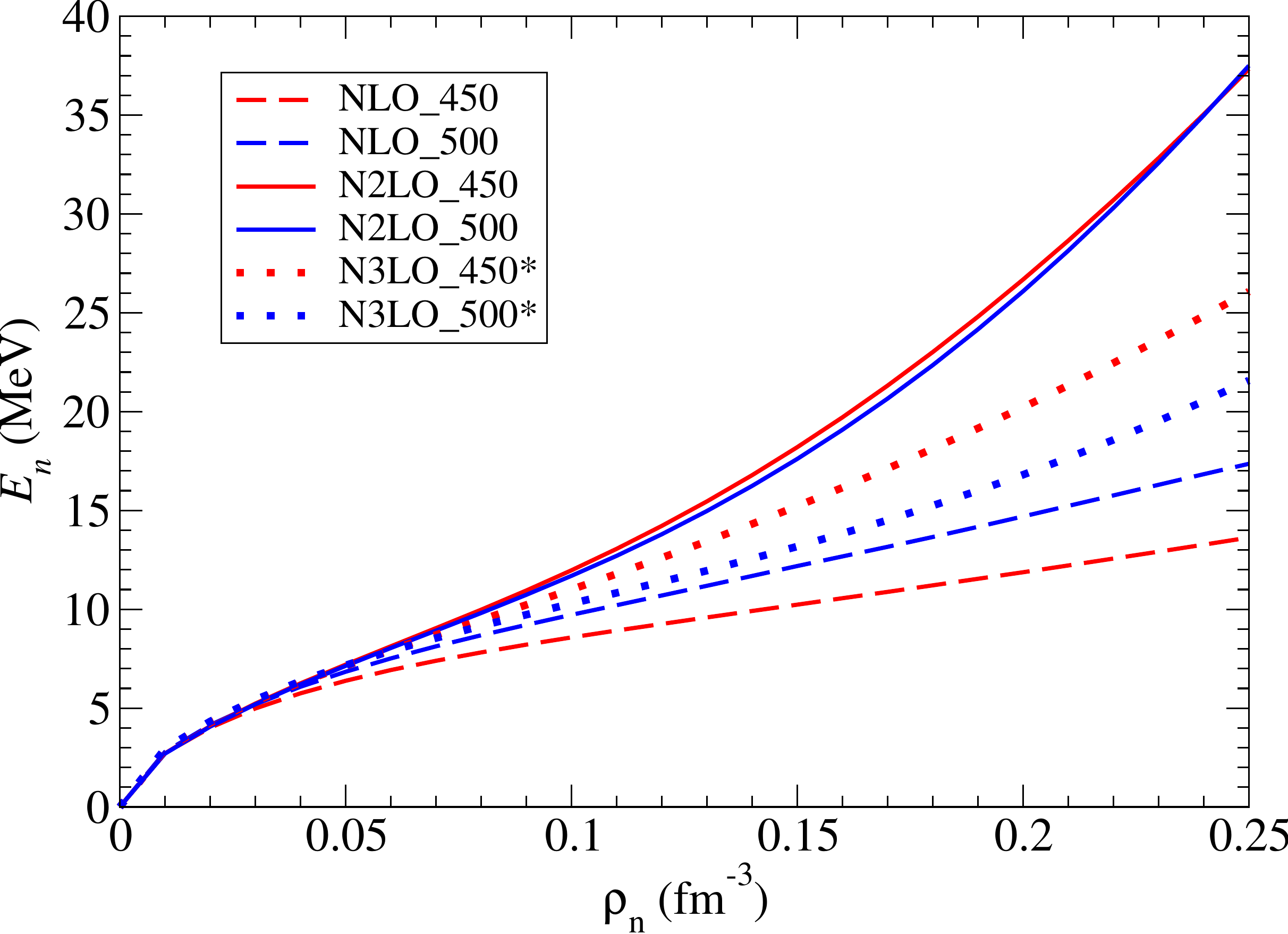}}
\caption{(Color online) Order-by-order convergence pattern of the neutron matter equation of state 
from NLO, N2LO, and N3LO* chiral two- and three-body potentials at third-order in
many-body perturbation theory with self-consistent second-order single-particle
energies. Two momentum-space cutoffs $\Lambda =$ 450\, MeV and 500\,MeV were chosen.} 
\label{e_obo_10}
\end{figure}

We show in Fig.\ \ref{e_obo_10} the equation of state calculated at third order in 
perturbation theory from the NLO, N2LO, and N3LO* chiral potentials with two choices 
of the momentum-space regulating scale $\Lambda =$ 450\,MeV and 500\,MeV.
In all cases the low-energy constants in the chiral two-body force are refitted \cite{marji13}
as a function of $\Lambda$ to $N\!N$-scattering
phase shifts and deuteron properties. As originally observed in 
Ref.\ \cite{sammarruca15} there is a large change from NLO to N2LO and also from N2LO
to N3LO*, indicating that neglected contributions may be a very significant source of theoretical
uncertainty. Comparing the ratio of differences $R^\Lambda_4 = 
(E_n^{(4)}-E_n^{(3)}) / (E_n^{(3)}-E_n^{(2)})$ for the two sets of chiral potentials, 
where $E_n^{(i)}$ is the neutron matter energy per particle at order $(q/\Lambda_\chi)^i$,
we find that $R_4^{450} \simeq 0.4$ and $R_4^{500} \simeq 0.8$ for all but the lowest densities.

\begin{figure}[t]
\centerline{\includegraphics[width=6cm,angle=270]{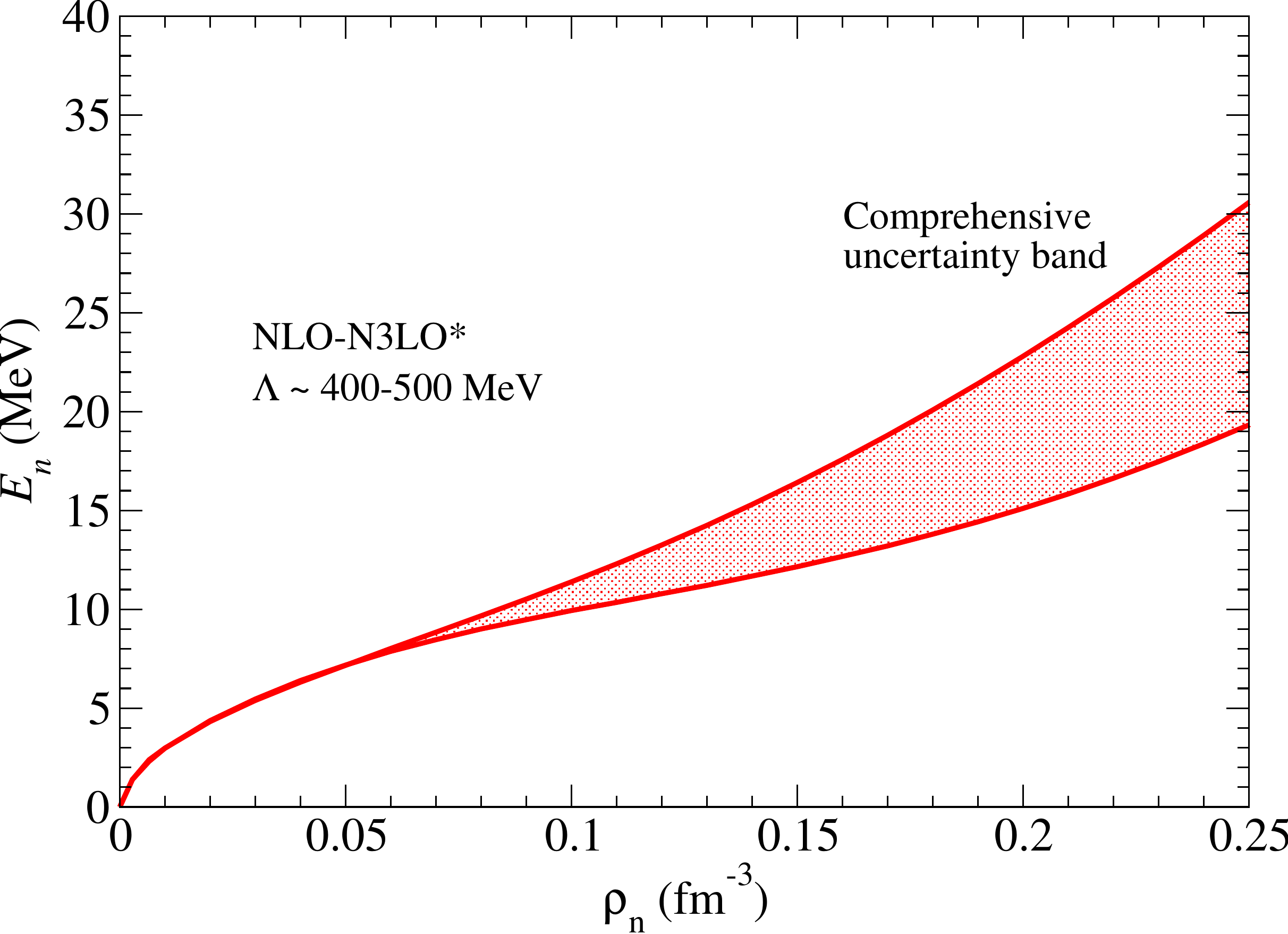}}
\caption{(Color online) Theoretical uncertainty estimate for the neutron matter equation of
state from chiral effective field theory, including errors arising from the convergence in 
many-body perturbation theory, missing terms in the chiral expansion, and choice of
resolution scale.} 
\label{pnmerror}
\end{figure}

In Fig.\ \ref{pnmerror} we show a comprehensive theoretical uncertainty estimate for the neutron
matter equation of state that accounts for errors due to truncations in many-body perturbation theory,
missing terms in the chiral effective field theory expansion, and the choice of resolution scale. 
The largest source of error in the present analysis is estimated to arise from missing higher-order 
contributions in chiral EFT. We have used the values of $R_4^{450}$ and $R_4^{500}$ to calculate
associated error bands on the N3LO* equations of state according to 
$E^\Lambda_{\rm N3LO*} \pm R_4^\Lambda (E_n^{(4)}-E_n^{(3)})$. In the case of the N3LO* chiral potential with
$\Lambda = 500$\,MeV, the lower band on the equation of state 
computed according to the above prescription would be well below even NLO results which 
include no repulsive three-body forces. We therefore limit the lower band of the uncertainty 
estimate by the scale dependence error, namely,
$E^{500}_{\rm N3LO*} - (E^{450}_{\rm N3LO*} -E^{500}_{\rm N3LO*})/2$.
In comparison to a recent calculation \cite{drischler16} of the neutron matter equation of state
and associated uncertainty estimate, our results exhibit a smaller theoretical error at low densities
but comparable uncertainties beyond $\rho = \rho_0$. The reduction in the low-density error
is directly attributed in the present calculation to the inclusion of third-order perturbative
contributions.

Theoretical predictions for the isospin-asymmetry energy and its density dependence 
can be extracted directly from our equations of state of symmetric nuclear matter and pure neutron 
matter. However, due to the large uncertainties in the symmetric matter equation of state, it is more 
reliable to instead expand about the known empirical saturation point at $\rho_0 =0.16$\,fm$^{-3}$ and 
$E(\rho_0) = -16$\,MeV. We consider the three neutron matter equations of state calculated
at third order in perturbation theory employing N3LO* chiral two- and three-body potentials. We
include as well the minimum and maximum on the uncertainty band shown in Fig.\,\ref{pnmerror}.
This gives a total of five neutron matter equations of state from which we extract the 
isospin-asymmetry energy at saturation density, $S_0 = S(\rho_0)$, and the associated slope parameter
\begin{equation}
\left .L = 3 \rho_0 \frac{\partial S(\rho)}{\partial \rho} \right |_{\rho_0}.
\end{equation}

\begin{figure}[t]
\centerline{\includegraphics[width=8cm,angle=270]{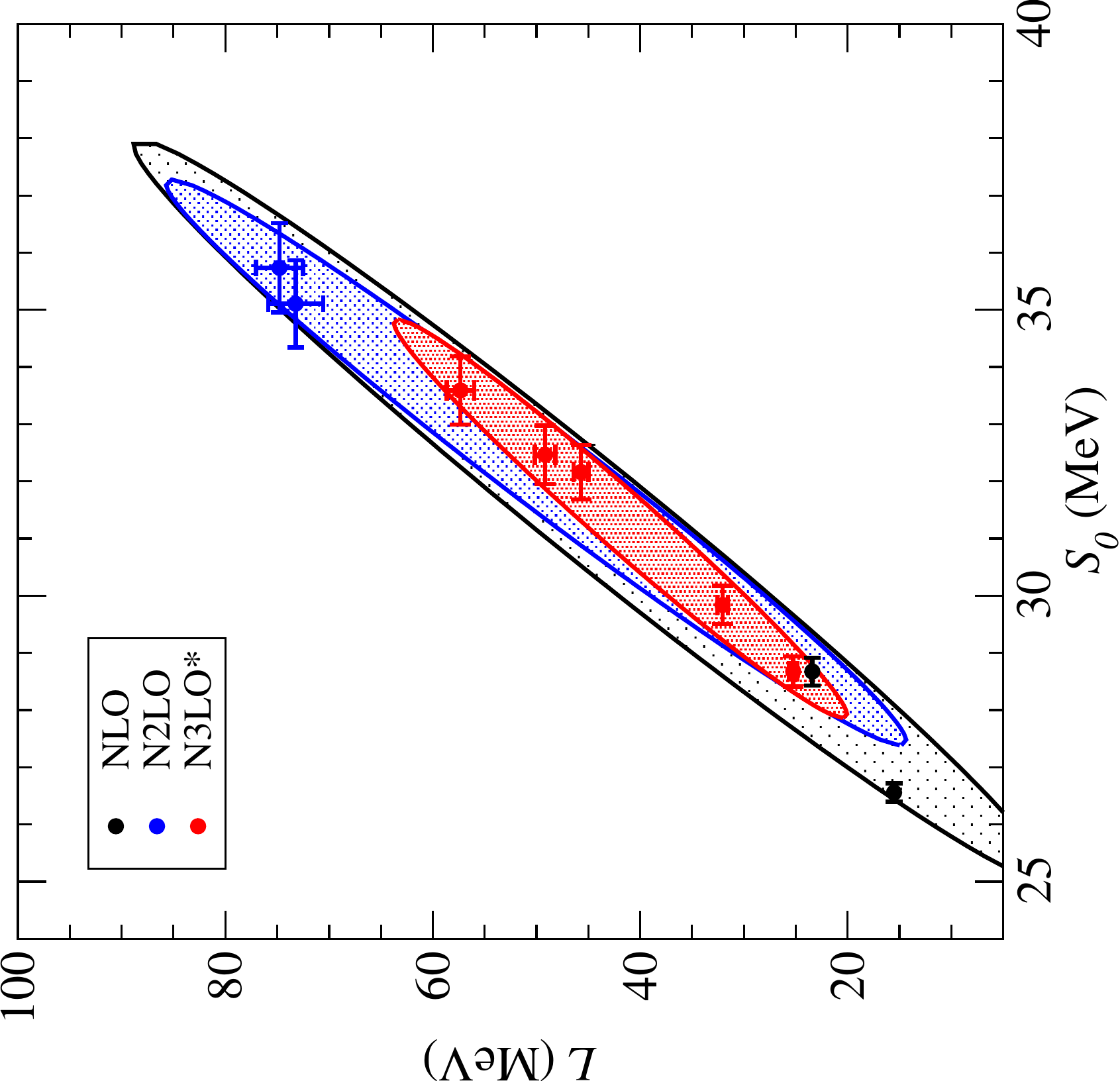}}
\caption{(Color online) 95\% confidence bands for the $S_0$ vs.\ $L$ correlation at order N3LO* from
chiral two- and three-nuclear forces (shown in red) and including also the N2LO (blue) and NLO 
(gray) values.} 
\label{svslbig}
\end{figure}

In Fig.\ \ref{svslbig} we show the correlation between $L$ and $S_0$ computed from 
NLO, N2LO, and N3LO* chiral two- and three-body forces at third order in many-body
perturbation theory. The error bars on individual points are obtained by 
varying the saturation density between $\rho_0 = 0.155$\,fm$^{-3}$ and 0.165\,fm$^{-3}$, keeping the
saturation energy fixed at $E(\rho_0) = -16$\,MeV. The two LO results from the chiral potentials with
$\Lambda = 450$\,MeV and 500\,MeV are shown in black and
give the lowest values of both $S_0$ and $L$ in the range $26<S_0<29$\,MeV and $15<L<25$\,MeV. 
The NLO results are shown in blue 
and give the largest values of the isospin-asymmetry energy and its slope parameter: 
$34<S_0<36$\,MeV and $70<L<80$\,MeV. Finally, the five equations of state at N3LO* give the
$S_0$ and $L$ values shown in red, which are in very good agreement with the results from 
previous microscopic calculations \cite{hebeler10prl,gandolfi12}.

In Fig.\ \ref{svslbig} we have also drawn $S_0$ vs.\ $L$ correlation ellipses at the 95\% confidence level
including only the N3LO* results (shown in red) as well as including the values of $S_0$ and $L$ 
from the NLO and N2LO equations of state. From the N3LO* correlation ellipse we infer a value of the
isospin-asymmetry energy in the range $28<S_0<35$\,MeV and slope parameter in the range
$20<L<65$\,MeV. The upper and lower data points in the N3LO* band come from including the 
uncertainty due to missing physics, which effectively introduces an additional error in the theoretical
prediction of the isospin-asymmetry energy on the order of $\Delta S_0 = \pm 2$\,MeV. For the
slope parameter, the effect of missing physics is to extend the theory prediction by about
$\Delta L = \pm 10$\,MeV. More conservative estimates on the $S$ vs.\ $L$ correlation are obtained
by replacing the upper and lower N3LO* points by those from the NLO and N2LO equations of state,
and the resulting correlation ellipses are shown in gray and blue in Fig.\ \ref{svslbig}.
The parameters associated with the three correlation ellipses are given in Table \ref{etab}. 
Remarkably the inclusion of NLO and N2LO equations of state modifies only slightly the 
inclination angle of the correlation ellipse, indicating a robust uncertainty estimate.

In passing we note that the analytical calculation of the third-order ring-diagrams from an $S$-wave contact 
interaction $V_{\text{ct}} = -{\pi \over M}[a_s+3a_t+(a_t-a_s)\vec \sigma_1\!\cdot\!\vec \sigma_2]$ provides
also a check on the validity of the (commonly used) quadratic approximation in the isospin-asymmetry $\delta = 
(\rho_n-\rho_p)/\rho$. Introducing a $pn$-mixed polarization function and expanding all occurring terms up 
to order $\delta^2$, one finds the following exact expression for the quadratic isospin-asymmetry energy:
\begin{eqnarray}
&& S_2(\rho)= {k_f^5\over \pi^4 M}\big(3.5124\, a_s^3 + 11.092\, a_s^2 a_t \nonumber\\ 
&& \qquad \quad  + 10.137\, a_s a_t^2 -5.1014\, a_t^3\big) \,.
\end{eqnarray} 
On the other hand the difference between the neutron matter and nuclear matter energy per particle 
(see Eqs.\ (13,14)) gives:   
\begin{eqnarray}
&& E_n(\rho)- E(\rho)= {k_f^5\over \pi^4 M}\big(3.6330\, a_s^3 + 9.4333\, a_s^2 a_t 
\nonumber \\ && \qquad \qquad \qquad \quad + 9.4333\, a_s a_t^2 - 5.2407\,a_t^3 \big) \,.
\end{eqnarray} 
One observes that the numerical coefficients of the cubic terms $a^3_{s,t}$ agree within $3\%$, whereas 
those of the interference terms are underestimated by $7\%$ and $15\%$ in the quadratic approximation. 
Moreover, when continuing the expansion in $\delta$ further one encounters a non-analytical term of the 
form $\delta^4 \ln|\delta|$. It has also been found in a second-order calculation with 
$V_{\text{ct}}$ in Ref.\ \cite{kaiser15}.

\setlength{\tabcolsep}{.035in}
\begin{table}[t]
\begin{tabular}{|c|c|c|c|c|c|} \hline
                         & $S_0 (\rm{MeV})$ & $L_0 (\rm{MeV})$ & $a (\rm{MeV})$ & $b (\rm{MeV})$ & $\tan \theta$ \\ \hline
    N3LO*         & $31.3$ & $41.9$ & $22.2$ & $0.64$ & $6.37$ \\ \hline
N2LO$-$N3LO* & $32.3$ & $50.0$ & $36.0$ & $0.86$ & $7.32$ \\ \hline
NLO$-$N3LO* & $31.5$ & $44.8$ & $44.5$ & $1.10$ & $6.91$ \\ \hline
\end{tabular}
\caption{Parameters of the $S$ vs.\ $L$ correlation ellipses (at 95\% confidence level) 
obtained at order N3LO* and
including also the NLO and N2LO points. The center values are labeled $S_0$ and $L_0$, 
the semi-major and semi-minor axes are labeled $a$ and $b$, and the inclination angle 
is labeled $\theta$.}
\label{etab}
\end{table}

\section{Summary and conclusions}

We have computed the equation of state of symmetric nuclear matter and pure neutron
matter including all  diagrams of many-body perturbation theory up to third-order with 
intermediate-state energies calculated self-consistently at second order. We have derived
semi-analytical results for the third-order particle-hole ring-diagrams from model-type interactions
that provide valuable benchmarks for numerical calculations based on a partial-wave
decomposition. We then employed realistic chiral two- and three-nucleon forces constructed at 
different orders in the chiral expansion together with a range of 
momentum-space cutoffs $\Lambda$ to compute the energy per particle of symmetric
matter and pure neutron matter, $E(\rho)$ and $E_n(\rho_n)$. The main motivation was to provide improved
theoretical uncertainty estimates on the isospin-asymmetry energy  $S_0$ at saturation density 
and its associated slope parameter $L$. We find that the convergence in 
many-body perturbation theory for the neutron matter equation of state is well under control 
at third order, and variations due to the choice of resolution scale are also relatively small
up to $\rho = 0.25$\,fm$^{-3}$. The largest theoretical uncertainty comes from higher-order
contributions in the chiral expansion, which we estimate by comparison of the equations of
state from NLO and N2LO chiral potentials. The derived correlation bands between $S_0$ and $L$ can
be used in updated global analyses of the density-dependent isospin-asymmetry energy $S(\rho)$.

\bibliographystyle{apsrev4-1}

%



\end{document}